\documentclass[12pt]{article}%
\usepackage{amsmath}%
\usepackage{amsfonts}%
\usepackage{amssymb}%
\usepackage{graphicx}
\providecommand{\U}[1]{\protect\rule{.1in}{.1in}}

\begin{document}
\bigskip\begin{titlepage}
\begin{flushright}
PUPT \\
hep-th/yymmnnn
\end{flushright}
\vspace{7 mm}
\begin{center}
\huge{ Kenneth Wilson in Moscow}
\end{center}
\vspace{10 mm}
\begin{center}
{\large
A.M.~Polyakov\\
}
\vspace{3mm}
Joseph Henry Laboratories\\
Princeton University\\
Princeton, New Jersey 08544
\end{center}
\vspace{7mm}
\begin{center}
{\large Abstract}
\end{center}
\noindent
I try to recreate the scientific atmosphere during Wilson's visits to Moscow in the '60-s and 70'-s.
\begin{flushleft}
January 2015
\end{flushleft}
\end{titlepage}\bigskip\ 

By the end of the sixties a few people , separated by the iron curtain, worked
on the theory of critical phenomena. The brilliant mind of Kenneth Wilson and
his visits to Moscow played a very important role in the subsequent
development. In this article I shall try to reconstruct the atmosphere of
these years . I have \ some "insider information " ( restricted to the Russian
side of the curtain ) which may be interesting .

It seems to me that the modern development of the subject started with the
work by Patashinskii and Pokrovsky \cite{pat1} .\ The rumors go that by the
end of the '50-s Landau already realized that his theory of phase transition
is incomplete and that the fluctuation corrections were important, but I have
never seen any written evidence of that and so my story begins with
\cite{pat1} .

It was an ambitious and complicated paper. The revolutionary assumption there
was the statement that in the Dyson equation $G^{-1}=G_{0}^{-1}-\Sigma\lbrack
G]$ the contribution from the free Green function ( the first term on the
right ) is negligible . Analogously, the bare contributions to the equations
for the vertex parts are also irrelevant. \ This assumption immediately leads
to \ "universality" -some small variations of the hamiltonian do not change
the critical behavior, since they perturb only the bare Green functions . Some
years later , Wilson introduced the concept of "irrelevant operators" which
generalizes and reformulates the Patashinski - Pokrovsky idea. The next step
in the paper was to guess that the equations are scale -invariant and to look
for the power- like solutions for the Green functions. Unfortunately, they
wanted too much and made some incorrect assumptions which fixed the critical
exponents. However , in '66 they quickly realized that and wrote another paper
\cite{pat2} which phenomenologically introduced scale invariance with
anomalous dimensions, this time in a completely correct way . It was unclear,
however,how this picture   was related to the QFT approach. The issue was
clarified in the different context ( Regge calculus ) by Gribov and Migdal
\cite{gribov} . They realized that with the correct treatment, the power- like
propagator is consistent with QFT for arbitrary exponents, which are
eventually determined by some bootstrap conditions. After that, in '67 -'68,
the papers \cite{pol1} and \cite{migdal1} applied these ideas to critical
phenomena. An interesting by-product of these works was the realization that
these phenomena are described by a relativistic field theory ( after Wick's rotation).

There was another , extremely important work , by Larkin and Khmelnitsky of '
\cite{larkin} \ . They have solved in '69 the four dimensional critical
theory, using the leading logs summation, equivalent to the Gell-Mann -Low
renormalization group. This work was just a half step from the $\varepsilon$ -
expansion (all one had to do was to replace the logarithm in the
Larkin-Khmelnitsky solution by a power, $\log p\Longrightarrow p^{\varepsilon
}/\varepsilon$), but this had to wait for two years when it was discovered by
Wilson and Fisher, together with the way to calculate the higher order corrections.

In '69 I tried to apply these ideas to deep inelastic scattering, realizing
that anomalous dimensions should break Bjorken's scaling in a multifractal
way. At that time I ordered in the public library the latest Phys. Rev.(which
usually took a couple of weeks to get) and experienced a strong shock ,
finding very similar ideas in the article by some Kenneth Wilson, the name
unknown to me. I still remember walking like a zombie through the center of
Moscow where the library was located.

And then, I think in the 1969 or 1970, Ken visited Moscow At that time Sasha
Migdal and I were passionately interested in what he was doing We had our own
approach based on the bootstrap ideas and Ken's renormalization group didn't
look promising to me. We spent hours with Ken, discussing these matters. \ His
approach at that time was based on the approximate recursion formula. Trying
to understand it , I derived it by some crude truncation of Feynman's
diagrams. Ken liked the derivation ( and generously included it in his later
review), but I thought it just showed that the recursion formula was too
primitive.However, later it helped Ken to develop a general approach to the
renormalization group and epsilon expansion.

In spite of our different "ideologies", I was very impressed by the power and
depth of Ken's arguments, and learned a lot of subtle things from our
discussions. One example was the operator product expansions (OPE\ ). In '69
they have been introduced in the various forms by Wilson \cite{wilson} , L
Kadanoff \cite{kad} and myself \cite{pol2} . While the general ideas were the
same in these three papers, the deepest version definitely belongs to Ken .
Namely, he traced the relation of the OPE to the canonical commutations
relation. It impressed me very much and I started to think that fields theory
in general should be defined by means of the OPE, the associativity of which
must restrict possible theories. This procedure is analogous to the
classification of simple Lie algebras (one of the most beautiful parts of
mathematics in my view) , but infinitely more subtle. These dreams became more
realistic with the '70 discovery of the conformal 3-point function, which made
the bootstrap equations concrete. I spent a lot of time trying the conformal
field theory (CFT) approach for gravity, but so far unsuccessfully. Ken was
not very enthusiastic about CFT. It was my impression (which may be wrong )
that he valued his \ version of renormalization group much more than the OPE
and was not much interested in their relations to each other. I had , and
still have , the opposite view and expect some big surprises in\ the
structures of higher dimensional field theories.

Speaking of field theories , I should add that in the 60's the high energy
theorists believed that field theory is a wrong way to approach Nature. Landau
had expressed this point of view a decade before, but some people became more
Catholic than the Pope.\ Not only the leading theorists in Russia were
sceptical about conformal field theory, but at the '70 Kiev Conference C.N.
Yang expressed strong disagreement with my comment on the relation between
critical phenomena and scale invariant QFT (this amusing exchange can be found
in the proceedings). It was heart warming for Sasha and me to see that Ken
Wilson did not share these prejudices, and that the idea that particle physics
and critical phenomena are related was as natural for him as it was for us.

In the seventies Wilson's methods , renormalization group and epsilon
expansion became tremendously popular and effective. They were easy to use in
numerical simulations ( this feature was very important for Ken ) , they also
gave a nice qualitative picture of a system. The terms like "UV fixed bpoint"
or "irrelevant operator ", introduced by Ken, became a part of the physics
dictionary. Still if we talk about exact analytic results, Wilson's
renormalization group is fully equivalent to the one by Gell-Mann and Low \ .
But I don't think that it bothered Ken. In our discussions he said something
like \ " Why should you care to get exact solutions ? After all, from the
computer point of view, the special functions are no different from any
expression which you can calculate with good precision. " ( I am not sure that
this was the exact wording, but\ I hope that the meaning of the phrase is
accurate). I disagreed, saying that if , say, a Bessel function appears in my
calculation, it unites\ my problem with the innumerable other theories. Sasha
Migdal at that time held the views very close to Ken's . Among many other
things, he later improved Ken's formulation of the renormalization group . Of
course , it is senseless to discuss who was right in this disagreement .

Next comes the theory of quark confinement and lattice gauge theory. This was
also a major development . For the first time the precise formulation of
confining gauge theory was given. Basically , Ken understood that what keeps
quarks together are the quantized Faraday flux lines, forming a string. He
then introduced the confinement criterion in terms of expectation values of
the phase factor, now called the Wilson loop . This led to many efforts to
derive confinement from the first principles. In the quasiabelian case (the so
called compact QED) this was indeed possible and Ken appreciated these
results. But the general problem of confinement is still unsolved . On the
other hand, the lattice gauge theory became an immensly popular and useful
tool for calculating physical properties of hadrons. The history of this
outstanding development was very nicely described by Ken himself
\cite{wilson1}. I can only add that in the ' 72 dissertation by Vadim
Berezinsky , the U(1) lattice gauge theory was explicitly written down. It
served as an inspiration for my non-abelian contribution to the subject,
mentioned by Ken.

At about the same time t' Hooft proposed 1/\ N expansion and conjectured that
the lines of the planar Feynman diagram will become dense and form something
like the string world sheet. It is important to distinguish these two
mechanisms ( which are often confused in the literature ). Electric flux lines
are not directly related to the propagator lines in Feynman's diagrams and in
QCD the diagrams do not become dense ( in the matrix models they do , but this
another story ). The modern gauge/strings duality is of the Wilson type.
Still, the large N expansion gives us control over the topology of the world
sheet, as well as a good phenomenological approximation.

In '79 in New York \ I was fortunate to have another scientific discussion
with Ken. I was anticipating it with great excitement, especially because I
had a number of new results , like a crude version of the non-critical string
theory and gauge/ strings duality , which , I hoped, should interest Ken.
Discussion with him always led to new insights.

Unfortunately, this time Ken was not interested at all. Our conversation was
fruitless . Perhaps I was unable to clearly communicate my ideas. And perhaps
Ken was changing his views on science. Be it as it may, but the resonance was
not there. It is sad to say that , but that was our last scientific interaction.

I would like to thank Boris Altshuler, Edouard Brezin, Igor Klebanov, Slava
Rychkov and Victor Yakhot for comments and advice. This work was partially
supported by the NSF\ grant PHY-1314198.

\end{document}